\documentstyle[12pt]{article} 
\begin{document}
\begin{center}
{\bf Radiative processes of nucleon interactions and
possible existence of exotic dibaryons}
\footnote{This work was supported by
the RFBR, grants No. 96-02-197147 and 96-15-96423 }\\.
{\bf S.B.Gerasimov \\
Bogoliubov Laboratory of Theoretical Physics, JINR,Dubna } \\
\end{center}
\begin{abstract}
The cross section of the reaction $pp \to pp \gamma \gamma$ is estimated
and the exploration possibilities of this reaction to probe for a possible
excitation of subnuclear degrees of freedom in two-nucleon systems,
in particular, the production and decay of the NN-decoupled dibaryon
resonances, are briefly discussed.Some arguments for and implications
of the intermediate dibaryon resonance excitation are inferred from
preliminary data on the two-photon yield and energy distribution
observed by the DIBAR2$\gamma$ - Collaboration at the kinetic energy
of the initial proton $T_{lab} \simeq 200 MeV$.
\end{abstract}

\section{Introduction}
In this report,we are addressing still unexplored reaction of two-photon
production in nucleon-nucleon interactions.Our immediate motivation to start
the discussion of this reaction is the experiment in progress at JINR
aimed at observing $2\gamma$-emission in $pp$-interaction below the pion
threshold and thereby to probe a possible existence of exotic dibaryon
resonances in the corresponding mass-range.We believe, however, that
these reactions merit dedicated studies on their own right.In particular,
they could contribute to our understanding of dynamics of two-current
processes with few-nucleon systems, like the Compton scattering on nuclei,
and to serve as a source of information on the role of subnucleonic degrees
of freedom in different kinematic conditions.The double (in general,
multiple) bremsstrahlung of photons by nucleons may be relevant to determine
the space-time extension of the emittance region, arising during
nucleus-nucleus collisions, via the study of the intensity
(or the Hanbury-Brown and Twiss ) correlations of photons, registered in
coincidence.Evidently, the observation of the two-photon emission is
interesting but very difficult experimentally, the main difficulties
being the low production cross section and requirements of good resolution
and an effective discrimination of the background.A few examples
illustrating this statement can be listed.For two-photon emission in
electron-atom collisions,the considerable efforts and time were needed
after the first claims \cite{Alt85} of observation of this process, to bring
subsequent measurements \cite{Liu93} into a qualitative agreement with
existing calculations \cite{Sm77}. In the area of low energy nuclear
and particle physics, two-photon decays of some levels in heavy
$({}^{131}Xe, {}^{137}Ba)$ \cite{Alv60} nuclei are known in the case
of competition with the one-photon $M4$- transitions, the observed
branching ratios being of the order $O(10^{-3})$. In $2\gamma$-decay of the
$0^{+}(3.35 MeV)$-excited state in ${}^{40}Ca$ \cite{Bear73},
where one-photon transitions are forbidden, and in the two-photon
$\pi^{-}$-absorption in nuclei \cite{Maz79} the relative
probability of $2\gamma$-emission is of the order $O(10^{-4})$. Taking these
branching ratios as a typical order-of-magnitude estimation for
the two-photon emission one can expect a very favorable background
situation while looking for resonance enhancement effects in the reaction
$pp \to pp \gamma \gamma$ with even a weak resonance signal.
The explicit estimation of both the non-resonance and resonance
mechanism contribution to cross sections will be given in Section 3.\\

\section{Experimental Study Motivations and Data}
In this section some results are reproduced of the first study of the
reaction $pp \to pp \gamma \gamma$ at the proton energy
$\sim 200 MeV$ \cite{DIB96}.

We believe that in addition to the ordinary bremsstrahlung,
$pp \to pp \gamma$, the two-photon production process also
deserves  a special investigation as a source, maybe, of a unique
information on some important aspects of hadrodynamics, underlying
the NN-interaction. In particular, the primary goal in this study
was to probe possible existence of exotic narrow dibaryon resonances
that could escape distinct observation in other reactions used
earlier for the same aim.

Existence of narrow states with a baryon number $B=2$ was
considered in several QCD-inspired models \cite{QCDM,Kondr87,Kopel95}
and in alternative standard models of the NN-interaction\cite{Pokrov}.
However, all available predictions for their masses and widths are
model-dependent and, therefore, cannot be treated as reliable yet.
The experimental situation has been somewhat confused.
Although a number of claims were made for the observation of
narrow structures \cite{Bes90,BublCh,Eldet,Tat92}, some of them were
not confirmed in later experiments \cite{Alesh92}.
Most of the dedicated experiments performed so far were aimed at looking
for NN-coupled dibaryon resonances.
Meanwhile, one might consider some processes leading to formation
of dibaryons with quantum numbers for which the direct decay $^2B \to $ NN
is either forbidden by the Pauli principle, or strongly suppressed by
the isospin selection rule (NN-decoupled dibaryons).
In this respect the process $pp \to \gamma {^2B} \to pp \gamma \gamma$
has unique possibilities of searching for and investigating narrow
NN-decoupled dibaryon resonances with masses below the pion production
threshold\cite{GerKh93,EGK95}.

The method of searching for these  narrow dibaryon
resonances is based on the measurement of the photon
energy distribution in the $pp \to pp \gamma \gamma$ reaction by detecting both
photons in coincidence. The narrow dibaryons, if they exist, should be seen
as sharp $\gamma$-lines against a smooth background due to
the photons from the radiative resonance decays($^2B \to \gamma pp$)
and double pp-bremsstrahlung with an anticipated good signal-to-background
ratio. The position of this line depends on the energy of the incident
proton and the resonance mass M$_B$. Its width is determined by the total
width of resonance $\Gamma_{tot}$ and energy resolutions of the experimental
setup. In Ref.\cite{Khr1} the first results were reported of searching for
the narrow NN-decoupled dibaryons in the process
$pp \to \gamma {^2B} \to pp \gamma \gamma$ at the proton energy
$\sim 200 MeV$. Measurements of the $\gamma$-ray energy spectra of this
reaction showed a well noticeable peak at the energy $E_\gamma \sim$ 47 MeV,
which was interpreted as evidence for the narrow dibaryon resonance
with a mass $\sim$1917 MeV.

Unfortunately, the statistics of that experiment was insufficient to draw
any firm conclusion on existence of the narrow dibaryon.

In what follows some results are reproduced from Ref.\cite{DIB96} on new
measurements of the photon energy spectra for the process:
\begin{equation}
pp \to \gamma ~ ^2B \to \gamma \gamma pp
\end{equation}
The experiment was performed with a proton beam from the JINR
phasotron with the proton energy 198 MeV and the energy spread
about 1.5\%. The experimental setup includes a liquid hydrogen target
and two detectors of $\gamma$-quanta placed on either side of the beam
to detect backward emitted photons at angles of 111$^0$ and 240$^0$,
respectively. The solid angles covered by photon detectors were 35 msr
and 70 msr, respectively.
The energy spectra of $\gamma$-rays were measured in the energy range
from 10 to 100 MeV.In the spectrum obtained as a result of
subtraction of the empty target contribution a structure at energy near
42 MeV is clearly seen.A fit to a Gaussian gave the energy of the
observed peak 42.0$\pm$4.5 MeV with  width(FWHM) 26.3$\pm$4.0 MeV.
If one assumes that a narrow dibaryon resonance is responsible for this peak,
then one can reconstruct the corresponding dibaryon mass distribution\cite{EGK95}.
This distribution has a maximum at $M_B \sim$ 1923 MeV.When fitted with
a Gaussian, the distribution gave a fitted mass 1923.5$\pm$4.5 MeV and
a width 31.3$\pm$5.0 MeV.\\
To conclude this section, the $\gamma$-ray energy spectrum for the
$pp \to pp \gamma \gamma$ reaction at the proton energy 198 MeV has
been measured in Ref.\cite{DIB96}. A distinct enhancement at the photon
energy about 42 MeV was observed in measured energy spectrum.
It can be interpreted as a signal due to the narrow exotic
dibaryon $^2B$ formation and decay in the $pp \to \gamma ^2B \to pp \gamma
\gamma$ processes. Distribution of the dibaryon mass obtained under this
assumption shows a narrow peak with mass $M_B$=1923.5$\pm$4.5 MeV and width
FWHM=31.3$\pm$5.0 MeV.
The statistical significance of this peak exceeds 8$\sigma$. The results
presented in Ref.\cite{DIB96} are in agreement with the previous
ones \cite{Khr1}.

\section{A Model Estimation of Effect and Background.}

In two decades after the first attempts of theoretical description of the
six-quark (or dibaryon) states\cite{Jaf77} the whole situation in theory
remains obscure. The lattice QCD approaches to the multiquark,e.g. the
$\bar q^2 q^2$- states\cite{Green93} were instrumental to-date only in
the denying of some simple pairwise forms of the $qq$-interaction,
which were guessed and used in the potential and cluster models of
multiquark ( mainly, six-quark) states.Therefore we adopt the explicitly
phenomenological approach in our estimations. Having in mind the
completeness of colourless hadron states, we shall estimate the probability
of the radiative transition $d_{1}(IJ^{P}=11^{+}) \to \gamma pp$ or
the inverse reaction $pp \to \gamma d_{1}$ as a two-step process, where the
presumably lowest pp-decoupled state with the $J^P=1^{+}$, tentatively called
$d_{1}$, is coupled with the initial or final hadron states through the
intermediate $N \Delta$-state with the same quantum numbers.The $"\Delta"$ -
symbol may be referred also to the virtual $\pi N$-complex with quantum
numbers of the $\Delta(1232)$-resonance but a different invariant mass.
The $d_{1}\Delta N$-vertex is described by a simple form of the
quasi-two-body wave function, for which the Hulthen-type radial
dependence was chosen by analogy with the deuteron radial wave function:
\begin{equation}
R(r) = N\frac{1}{r}exp(-\alpha r)(1 - exp(-\beta(r-r_c)))
\end{equation}
where N is the normalization constant,$\alpha=\sqrt{2M_{red}\varepsilon},
\varepsilon=M+M_{\Delta}-M_{d_1},
M_{red}^{-1}=M^{-1}+M_{\Delta}^{-1},\beta=5.4 fm^{-1}, r_{c}=.5 fm$ and
$R(r)=0$ for $r \leq r_{c}$ is understood. The second factor in Eq.(2),
representing the behavior of wave function in the "interior" region
outside the hard core with the radius of $r_c=.5 fm$ is taken quite similar
to the deuteron case.Taking $M_{d_1} \simeq 1920 MeV$ for granted,
the transition magnetic moment $\mu(p \Delta^{+})=2\sqrt{2}/3 \mu(p)$
according to the $SU(6)$-symmetry and plane waves for initial protons,
we get an estimation
\begin{equation}
(\frac {d\sigma}{d\Omega_{\gamma}})_{c.m.} \simeq \frac {\alpha(W-M_{d_1})^3(\mu(p)I(q))^2}{9Mq} \simeq 40 \frac{nb}{sr}
\end{equation}
$$ I(q) = \int_{r_c}^{\infty}dr r^2R(r)\frac{\sin(qr)}{qr}$$
where $\alpha=1/137, \mu(p)=2.79,q \simeq \sqrt{MT_{lab}/2}$,
$M$ is the mass of the proton.
It turns out justified to neglect the retardation corrections,
i.e. we are using the long-wave approximation for the matrix element of
magnetic-dipole transition.Further,the result does not depend strongly on
variation of the "effective" mass $M_{\Delta}$ from $M+m_{\pi}$ to
$M_{\Delta}=1232 MeV$.With the integrated luminosity $L=10^{37} cm^{-2}$,
the data gave about 130 events after integration over energies of both
photons in the interval  $10 \leq \omega_{i} \leq 100 MeV$.Assuming the
spherical-symmetric distribution of photons, we obtain an estimate
of the cross section for one of photons to be registered
in the element of solid angle
\begin{equation} \frac
{d\sigma}{d\Omega_{\gamma}} \simeq 4\pi(\frac {d\sigma}{d\Omega_{\gamma_1}
d\Omega_{\gamma_2}})_{exp} \simeq 65\frac{nb}{sr}
\end{equation}
By definition, this is the quantity which should be confronted
with Eq.(3) and we estimate the result of this comparison as a reasonable
one.\\
We turn now to the usual non-resonance double-bremsstrahlung reaction.
The most reliable would be the calculation based on the NN-potential
models taking into account off-shell effects in nucleon-nucleon interactions
explicitly.  This is, however, an extremely complex task requiring much
of numerical computations.To perform the exploratory calculation, we make use of the following
simplification. In the "suspected" resonance region all photons have energy
$\sim 40$ MeV at which, in one-photon bremsstrahlung reaction, the radiation
by magnetic moments of interacting protons becomes to dominate over the
non-spin-dependent convection current contributions.We adopt this magnetic
transition dominance also for the two-photon emission in our specific energy
region.Therefore two photon lines are attached in all possible ways to the
external proton lines of the corresponding Feynman graphs.A set of such
diagrams represents the amplitude of the reaction considered.The most drastic
approximation is the use of a model construction $T_{av}$ instead of the
complicated off-shell T-matrix of NN-scattering entering into the Feynman
diagrams. The $T_{av}$ is assumed to be the spin- and angle - averaged
quantity, defined through the total pp-cross section:
\begin{equation}
\sigma_{c.m.}^{pp}(q)=\frac{1}{2}\int d\Omega_q \frac{d\sigma_{el}^{pp}}{d\Omega_q}=\frac{(WT_{av}(q))^2}{32\pi}
\end{equation}
where we keep only the energy dependence of cross sections according to
the following prescription. If both photons are radiated by the initial
(or final)
protons, then the total cross section is taken at the invariant energy
$W_f=\sqrt{s_f}=\sqrt{(p_1+p_2-k_1-k_2)^2}$ or $W_i=\sqrt{s_i}=\sqrt{(p_1+p_2)^2}$,
$p_j(k_j)$ are the 4-momenta of corresponding protons(photons), $j=1,2$..
If one of photons is radiated by initial protons and another by
final ones,we propose to use the average value $\bar W=1/2(W_1+W_2)$, where
$W_j=\sqrt{(p_1+p_2-k_j)^2}, j=1,2$. In calculation, we are using
the pp-cross sections from available compilation thereof, except at very low
energies, where the effective range parametrization with account of
the Coulomb interaction \cite{Slo68} is taken. We refrain from writing
down all standard phase space and normalization factors and cite only
the result obtained after integration over energies of photons in the
interval $10 \leq \omega_j \leq 100 MeV$,the angle between photon directions
being $\vartheta=130^{0}$:
\begin{equation}
(\frac {d\sigma^{2\gamma}}{d\Omega_{\gamma_1} d\Omega_{\gamma_2}})_{nonres} \simeq
\frac{\alpha^2 \mu_{p}^4}{128\pi^5M^2} \int d\omega_1 d\omega_2
\omega_1 \omega_2 (T_{av}(s_f) + T_{av}(s_i)^2 R(\omega_{1,2};\vartheta)
P(\omega_{1,2};\vartheta )\simeq 1.3\frac{pb}{sr^2}
\end{equation}
$$ R(\omega_{1,2};\vartheta) = 5a + 3b + (a - b)\cos^2 \vartheta $$
$$ a=(1 - 2T_{av}(\bar s)/(T_{av}(s_f)+T_{av}(s_i)))^2 ,
b=((\omega_2 - \omega_1)/(\omega_2 + \omega_1))^2 $$ \\
$$ P(\omega_{1,2};\vartheta ) = \sqrt {1- \frac {2(\omega_1+\omega_2)}{T_{lab}}-
\frac {\omega_1^2 + \omega_2^2 + 2\omega_1\omega_2 \cos \vartheta}{2MT_{lab}}} $$
\begin{equation}
\frac {d\sigma_{nonres}^{2\gamma}}{d\sigma_{exp}^{2\gamma}} < O(10^{-3})
\end{equation}
The estimation obtained as a level of the expected physical background
is seen to be much lower than the effect recorded, and this finding
reenforces the probability of its interpretation as due to the NN-decoupled
dibaryon resonance excitation with a subsequent dominant radiative decay.
We notice that the transition into the final state where two protons form
the singlet $^{1}S_{0}(0^{+},T=1)$ - state plays very significant role
in the $pp2\gamma$-reaction.It is worth to make a few qualitative remarks
about role of this "singlet level" in other bremsstrahlung reactions.
It is easy to see that the static $E1$- and $M1$- transitions into
the $^{1}S_{0}(pp)$ - state are absent in the $pp\gamma$-reaction, while
the $E2$-transition is small $\sim$ $O((p/m_{N})^4)$. Such the suppressing
factors are absent for the $np\gamma$-reaction. Therefore in the $np\gamma$
-reaction and especially in the $np 2\gamma$-reaction the transition to
the $^{1}S_{0}(np)$ - state should be much more important and accessible for
investigation.One can expect also a significant role of the meson exchange
currents (MEC) in the radiative $np$-reactions. In particular, the study
of the $np2\gamma$-reaction seems to be especially interesting as the check
of the MEC in different kinematic conditions: the $0^{+}-0^{+}$-transition
is possible in this case which is most sensitive to the short-range nucleon
interaction.Furthermore, the knowledge of this reaction is necessary for
correct interpretation of the H-B-T-type experiments dealing with
the $2\gamma$-interferometry in the nucleus-nucleus reactions
$ AA \to 2 \gamma X$ at energies below the pion threshold.

\section{Concluding Remarks}
We conclude with a few remarks.\\
1. Among theoretical models predicting dibaryon resonances with different masses
there is one giving the state with the $IJ^P=11^{+}$ and the mass value
$(\sim 1940 MeV)$ surprisingly close to the value $(\simeq 1920 MeV)$
extracted from the observed maximum of the $pp \to pp2\gamma$-reaction.
This is the chiral soliton model applied to the sector with the baryon
number $B=2$ \cite{Kopel95}.  The theoretical uncertainty at
the level of $\pm 30 MeV$ might be taken here because the model gives this
numerical (unrealistic) value for the mass difference of the deuteron
and the singlet level.However the cited radiative width of the order
$\sim O(eV)$ looks much too low.\\
2. Continuation of experiment in different kinematic conditions and,
hopefully, with an improved photon energy resolution is of undoubted importance
because the reliable discovery of even a single exotic multiquark state
would be immensely important and would lead to far reaching consequences for
the development of low-energy QCD in the domain both of hadron and nuclear
physics.\\
3. Closely related processes such as the (double) radiative pion-capture or
radiative muon-capture in deuterium mesoatoms would be very helpful not
only as different area of checking the very existence of exotic dibaryons,
but also as a means to discriminate between possible values of their isospin.
In particular, one can anticipate the spectacular anomalous high
branching ratio for the radiative muon-capture in deuterium $\mu$-mesoatoms
due to excitation in the intermediate state of the sufficiently low-lying
$NN$-decoupled isovector dibaryon(s), like one shown up in the
$2\gamma$-production, Ref.\cite{DIB96}, but excited this time by the charged
weak current.
In heavier nuclei this effect may be absent due to a more probable
radiationless decay of virtual dibaryon resonances in the nuclear field of
nearby nucleons.These possibilities deserve more detailed elaboration
and exposition that will be done elsewhere.

\section{Acknowledgements}
The author is much indebted to A.S.Khrykin for collaboration on
different parts of this report.
This work was supported in part by
the Russian Foundation for Fundamental Researches, grants No. 96-15-96423 and
96-02-19147.

\end{document}